# STEV: A Stabilized Explicit Variable-Load Solver with Machine Learning Acceleration for the Rapid Solution of Stiff Chemical Kinetics

K. Buchheit[a,b], O. Owoyele[a,b], T. Jordan[a], D.T. Van Essendelft[a]

[a] National Energy Technology Laboratory, Morgantown, WV 26505, USA

[b] Oak Ridge Institute for Science and Education, Oak Ridge, TN 37831, USA

## Abstract

In this study, a fast and stable machine-learned hybrid algorithm implemented in TensorFlow for the integration of stiff chemical kinetics is introduced. Numerical solutions to differential equations are at the core of computational fluid dynamics calculations. As the size and complexity of the simulations grow, so does the need for computational power and time. Solving the equations in parallel can dramatically reduce the time to solution. While traditionally done on CPU, unlocking the massive number of computational cores on GPUs is highly desirable. Many efforts have been made to implement stiff chemistry solvers on GPUs but have not been highly successful because of the logical divergence in traditional stiff algorithms like CVODE or LSODA. Because of these constrains, a novel Explicit Stabilized Variable-load (STEV) solver has been developed. In the STEV solver, overstepping due to the relatively large time steps is prevented by introducing limits to the maximum changes of chemical species per time step. Additionally, to prevent oscillations as the species concentrations approach steady state, a methodology to identify and dampen ringing via a discrete Fourier transform is introduced. In contrast to conventional explicit approaches, a variable-load approach is used here, where based on its thermodynamic state, each cell in the computational domain is advanced with its unique time step. This approach allows cells to be integrated simultaneously while maintaining warp convergence but finish at different iterations and be removed from the workload. To improve the computational performance of the introduced solver, specific thermodynamic quantities of interest were estimated using shallow neural networks in place of polynomial fits, leading to an additional 10% savings in clock time with minimal training and implementation requirements. While the complexity of these particular machine learning models is not high by modern standards, the impact on computational efficiency should not be ignored. Further, these gains will only increase as new generations of machine learning specific hardware become available. For example, it is estimated that if the machine learning portions of the solver were transitioned to the Volta architecture that the time savings would jump from 10% to as much as 28% due to the use of the new tensor core architecture. The results obtained from the new solver are compared to traditional LSODE solver used in the Multiphase Flow with Interphase eXchanges (MFiX) Computational Fluid Dynamics (CFD) code, showing a dramatic decrease in total chemistry solution time (over 200 times for single-phase and multiphase flows) while maintaining a similar degree of accuracy. The results also show that larger savings in computational time are achieved as the number of cells per node is increased, indicating the potential applicability of STEV to problems with larger cell count per node as compared to implicit-based solvers.

# 1. Introduction

Computational Fluid Dynamics (CFD) provides an excellent means to simulate a real system and streamline the design process by building predictive numerical models, obviating the need for a cost prohibitive process of testing several physical prototypes. In a reacting system, the evolution of species in space and time needs to be accounted for, in addition to the existing mass, energy, and momentum equations. Solving these equations can dramatically increase the simulation times because of the rapid changes in concentration that can occur through the chemical source terms. Furthermore, in many cases, chemical species have very different time scales over which the evolution in time occurs, leading to a stiff set of equations. In the case of stiff sets of chemical reactions, the simulation time could be completely dominated by the solving of the ordinary differential equations (ODEs) tied to the species rates. This study aims to improve the time-to-solution of the chemical rate integration by implementing a novel Stabilized, Explicit, Variable-load (STEV) solver method on both CPU and GPU by merging TensorFlow with NETL's in house CFD code, Multiphase Flow with Interphase eXchanges (MFiX).[1] This work represents, to the best of the authors' knowledge, the first effort to integrate machine learning (ML) libraries with a CFD solver.

TensorFlow is a static graph machine learning framework developed by Google.[2] It is now a flexible platform for a range of tasks that utilizes dataflow programming. The static graph methods in TensorFlow allows it to run an algorithm on multiple operating devices, such as CPUs, GPUs, or TPUs. This saves development time and allows for ease of deployment without needing to specify instructions for how the program should run on a given piece of hardware. Further, once the graph is compiled, there is no difference in operation speed between the TensorFlow generated operations and the equivalent machine specific language operations, unlike interpreted languages. Thus, an efficient graph will run as quickly as efficient compiled code and use the same math libraries. There are some minor startup costs to graph execution, but if enough work is done, the impact is exceptionally small. In addition several inference engines are being developed to both optimize the graph and more efficiently run the TensorFlow graph. Therefore, TensorFlow can be used as an exceptionally flexible and efficient math library to develop and deploy algorithms across a wide variety of the latest hardware accelerators for generalize computation as well as machine learning.

Most ODE solvers used in the solution of stiff chemistry utilize some form of implicit finite difference scheme such as the backward differentiation formula due to their ability to remain stable while taking relatively large time steps when compared to explicit methods. Implicit methods require the solution of non-linear (often system of) equations at each time step. Multi-core CPUs can handle the various logic elements present in an implicit solver algorithm as each thread can act independently. However, the same implicit methods implemented on the GPU have often ended up being slower than the CPU due to warp divergence.[3-6] This study aims to implement a novel method that can maintain warp convergence which will utilize the full parallelism of a GPU and solve stiff sets of ODEs. Furthermore, this study aims to accelerate the new method by hybridizing the solver with machine learning techniques.

# 2. The Stabilized, Explicit, Variable-load (STEV) Solver

It is well known that explicit integrators suffer from stability issues that drives the allowable time step to be small especially for stiff systems of equations. Thus, explicit methods have largely been abandoned in favor of implicit methods for stiff problems. Nonetheless, state-of-the-art implicit methods require a significant amount of logic within the algorithm to determine time step and order which results in poor performance on GPU's.[3-6] The STEV solver incorporates several stabilization methods to prevent overstepping and non-physical oscillations during integration. In addition, STEV allows each cell to advance with its own time step and cells are removed as they reach the desired integration time. In doing so, warp convergence is maintained which maximizes the efficiency on the GPUs.

In general, unstabilized explicit integrators will over step in concentration or time which will cause inaccurate integration, negative concentrations, and/or ringing. To extend the applicability of explicit methods into stiffer regions, targeted stabilization methods were developed which limit the negative effects. In this work, the problem of overstepping is mitigated by smart choices of the integration time step (see section 2.1). In addition to this, the negative concentration are minimized by species stabilization (see section 2.2), and the DFT stabilization (see section 2.3) is used to reduce the effects of ringing.

## 2.1 Variable Time Step Stabilization

The first unique characteristic of the STEV solver is that each cell can evolve on its own time step. While the accuracy of explicit methods is largely dependent on the time step, there is no need for the entire field to evolve at the same time step. What is required is that each cell integrates to the same end time.

The time step stabilization works to limit both the maximum time step and the maximum species consumption as seen in Eqs. (1) to (5) and graphically represented in Figure 1.

$$r_i^* = \max(-r_i, 1.0 \times 10^{-30}) \ \forall \ i \in (1 \ldots n) \tag{1}$$

$$y_i^* = \begin{cases} 1.0 & \text{if } y_i \ll 1 \times 10^{-20} \\ 0.9 y_i & \text{otherwise} \end{cases} \tag{2}$$

$$t^* = \min_i \left( \frac{\min(Y_{step}^{max}, y_i^*)}{r_i^*} \right) \tag{3}$$

$$\Delta t_{max} = t_f \cdot \delta^{max} \tag{4}$$

$$h = \min(t^*, \Delta t_{max}) \tag{5}$$

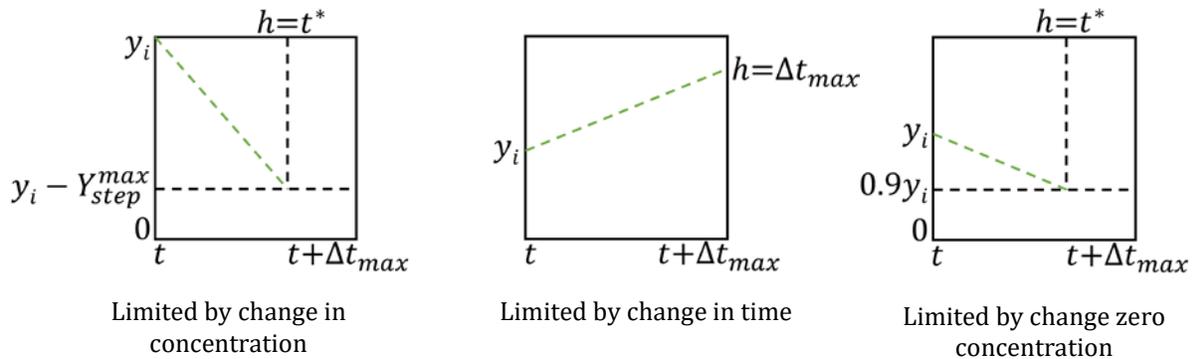

Limited by change in concentration · Limited by change in time · Limited by change zero concentration

*Figure 1: Graphical representation of time step stabilization*

In the above equations, $h$ is the time step calculated for each cell, $y_i$ is the species concentration in each cell, $r_i$ is the species rate of change in each cell, $Y_{step}^{max}$ is the maximum allowed species consumption at each step, $t_f$ is the total integration time, $\delta^{max}$ is a fraction which limits the maximum allowed time step, and $\Delta t_{max}$ is the maximum allowed time step.

The objective Eq (3) is to calculate allowable time step which will limit consumption to either a predefined maximum species consumption, $Y_{step}^{max}$, or ninety percent of the current species concentration. Species generation is not strictly controlled. However, by carefully controlling consumption and by the law of mass action, species generation is proportionally limited. Further, the species concentrations are appropriately controlled to be between zero and unity.

It is also worth noting that limiting the time step in this way effectively makes explicit methods usable for nearly infinitely stiff irreversible reactions if the time step can be represented within the numeric precision chosen. Moreover, it helps to increase the integration accuracy by dynamically limiting the changes when the rates are high in magnitude.

The objective of Eq (4) is to prevent over stepping when dealing with something like delayed ignition. In such cases, the temperature limits the stiff chemistry and it is necessary to control the time step to obtain the correct ignition delay. Typically, if ignition delay is important $t_f$ in the simulation is small and a fraction $\delta^{max}$ of between 0.001 and 0.01 is sufficient. Care should be taken in choosing this factor because it directly impacts the number of steps the solver must take to reach $t_f$ (i.e. the amount of time it takes for the solver to complete the integration).

## 2.2 Small Species Stabilization

To avoid negative concentrations, a term which artificially slows reactions in the limit of small concentrations was added to the chemical reaction rates. That is, $y_i$ is replaced with $\gamma(y_i)y_i$ in the rate equations where:

$$y_i^{**} = y_i \gamma(y_i) \quad (6)$$

$$\gamma(y_i) = \frac{y_i}{(\alpha + y_i)} \quad (7)$$

Here, $y_i^{**}$ is the effective species concentration, $\alpha$ is a species limit factor such that when $y_i \gg \alpha$, $\gamma(y_i)$ is unity and when $y_i \ll \alpha$, $\gamma(y_i)$ is zero. Typical values for $\alpha$ are near $1 \times 10^{-5}$. This stabilization technique works in conjunction with Eq (3) to smooth the transition of a stiff reactions when reagents approach a near zero concentration and lowers the number of steps needed to integrate in these cases. Typically, extremely low species concentrations are not important. If they are, lowering $\alpha$ or not using this technique is advisable.

## 2.3 Discrete Fourier Transform (DFT) Stabilization

The techniques in sections 2.1 and 2.2 help stabilize irreversible reactions and reactions with species concentrations near zero. However, they typically will not help in cases of stiff reversible reactions where the equilibrium point is not near zero species concentrations. In detailed combustion mechanisms, these reactions are common.

To help stabilize these rate expressions, a methodology based on the DFT of the change in reaction progress for the reversible reactions was developed. In this method the change in reaction progress for each cell is calculated and stored in an array as in Eq (8).

$$r_{store}^i = r_{rev} h \tag{8}$$

Here, $r_{store}^i$ is the i$^{th}$ stored value of the change in reaction progress, $r_{rev}$ is a reversible reaction rate, and h is the time step. Typically, only the thirteen most recent changes in reaction progress are kept for calculating the DFT. The magnitude of the DFT of $r_{store}^i$ is taken at each integrator time step and a ringing factor is calculated as,

$$D_{rev} = |DFT(r_{store}^i)| \tag{9}$$

$$f_{ring} = \frac{D_{rev}^0}{\max\left(D_{rev}^j\big|_{n<j<N}\right)} \tag{10}$$

Here, $D_{rev}$ is the magnitude of the DFT for each cell, $D_{rev}^j$ is the j$^{th}$ frequency component of the DFT, n denotes the start of high frequency components, N is the highest frequency component, and $f_{ring}$ is the ringing factor. If the low frequency term dominates $f_{ring}$, then it is near unity. As ringing occurs $f_{ring}$ drops. Further, as equilibrium is attained $D_{rev}^0$ approaches zero. Thus, the ringing equilibrium condition is detectable and controllable through Eqs (11) to (14).

$$rOF_0 = 1.0 \tag{11}$$

$$rOF^* = \begin{cases} 1.0/r_{fac} & \text{if } f_{ring} < Tol_{ring} \text{ and } D_{rev}^0 < Tol_{freq} \\ r_{fac} & \text{otherwise} \end{cases} \tag{12}$$

$$rOF_{t+1} = \min(rOF_t * rOF^*, 1.0) \tag{13}$$

$$r_{rev}^{eff} = r_{rev} * rOF_t \tag{14}$$

Here, $rOF$ denotes the rate oscillation factor and is limited to the range $0 < rOF \leq 1.0$. The subscript on $rOF$ denotes the time step while *0* is the initial time step, *t* is the current time step, and *t+1* is the next time step. $Tol_{ring}$ is a ringing tolerance which sets the allowable level of ringing and $Tol_{freq}$ is the low frequency tolerance which sets the threshold for equilibrium detection. Finally, $r_{fac}$ is the factor that determines how quickly to damp the reversible rate and $r_{rev}^{eff}$ is the effective rate used in the integration. $rOF$ is initialized as 1.0 and will not decline unless the reversible rate is both ringing and at equilibrium. If the reversible rate meets these criteria, it will quickly be dampened so that equilibrium is maintained, and the integration can continue without such small time steps. If the conditions change, the rate is quickly brought back up to full speed.

This method of controlling ringing was tested with the BFER mechanism for methane and found to be able to integrate these equations with good performance. The BFER mechanism is a six species, two step mechanism which was reduced from GRIMECH and fairly accurate with equivalence ratios below about 1.3.[7, 8] The important thing to note is that both reaction steps are exceptionally stiff and the CO oxidation reaction is reversible. See section 7.1 for more details.

That said, several detailed mechanisms like GRIMECH have many stiff, reversible, and competitive reactions. The performance of the STEV solver with DFT stabilization was degraded as the number of stiff, reversible, and competitive reactions increased and did not integrate GRIMECH well without using very small time steps. Performance improvements are still under development. However, the current version of the STEV solver is expected to perform well with almost all global combustion mechanisms and many analytically reduced mechanisms.

### 2.4 Variable Loading

Because each cell can evolve at its own time step, each cell will reach the final integration time at a different iteration. To deal with this additional complexity, a variable is initialized that contains the initial thermochemical state (species concentrations, temperature, and pressure). As cells reach the final integration time, the end thermochemical state is copied to the variable and the cell is removed from further calculation. In this way, the solver only works on unfinished cells during every iteration and maintains a logical coherence that results in no thread divergence. The only thing that changes is the "load" which is the number of cells that are being integrated at any given iteration. The STEV solver has the characteristic that for the first $1/\delta^{max}$ iterations, every cell is active. After that number of iterations, cells can drop out leaving only cells with stiff conditions. For computational fluid dynamics, this characteristic is extremely valuable as truly stiff conditions exist for a small portion of the simulation where there is a mix of oxidants and fuels and high temperatures. In this way, the STEV solver works only as hard as conditions in the simulation demand and can operate with near ideal parallelization.

## 3. The Machine Learning Accelerated STEV (MLA-STEV) Solver

Stiff chemistry solvers fall into the class of Initial Value Problems (IVPs) while most other problems in CFD are Boundary Value Problems (BVPs). This distinction turned out to be quite important in efforts to hybridize the STEV solver with Machine Learning (ML). The problem with using ML to predict the integration of the thermodynamic state is that the accuracy of the predicted state (while good by ML standards) is not sufficient when the thermodynamic state is repeatedly fed into the ML model as is done while running CFD. Multiple studies have focused on training neural networks for thermochemical integration.[9, 10] Significant accuracy gains were obtained relative to previous work with ML predictions of thermochemical states in this study. However, even with relative prediction errors reaching as low as $1 \times 10^{-5}$, the solution temperature in CFD slowly diverged from a physically meaningful value.

While it is true that evaluating the ML based predictions are exceptionally quick, the nature of the IVP does not allow for efficient correction of the ML prediction because any correction methodology involves either: 1) integration from the initial/current condition or 2) a large number ML predictions to form a time field as an initial guess to formulate the solution method as a nonlinear BVP. There is no point in correcting via 1) as it is just as much computational effort as integration by standard IVP solution methods. The computational efficiency of the ML predictions in 2) is diluted by the number of samples needed to establish the initial field and the computational effort of the nonlinear BVP solver. Further, is not possible to know how many samples are needed to maintain stability nonlinear BVP solver a priori.

Thus, ML acceleration was limited to indirect property calculations for thermodynamics as prediction errors were lower overall and errors had a lower impact on simulation results. The adiabatic temperature dependence in the system of ODEs requires the specie thermodynamic properties of enthalpy and heat capacity to be estimated. A common method of computational estimation involves the use of the NASA 7 polynomials which follow the forms in Eqs (15) and (16).

$$\frac{C_P}{R} = a_1 + a_2 T + a_3 T^2 + a_4 T^3 + a_5 T^4 \tag{15}$$

$$\frac{H}{RT} = a_1 + \frac{a_2 T}{2} + \frac{a_3 T^2}{3} + \frac{a_4 T^3}{4} + \frac{a_5 T^4}{5} + \frac{a_6}{T} \tag{16}$$

Where $C_P$ is the constant pressure heat capacity, $H$ is enthalpy, the ideal gas constant $R$, $T$ is the species temperature, and $a_i$ is the $i$ coefficient in the series. The thermodynamic properties are also represented across a temperature range with at least one split temperature, $T_{low} < T_{split} < T_{high}$, with a set of coefficients for each range. For calculation at run time, vectors of temperature need to be evaluated logically to determine where the temperature of each individual reacting cell falls into these ranges to use the correct coefficients. Enthalpy and heat capacity are smooth and continuous functions for each specie that is not undergoing a phase transition which makes it simple to estimate by a neural network. A single Artificial Neural Network (ANN) was trained to estimate the thermodynamic properties for all phases and all species simultaneously.

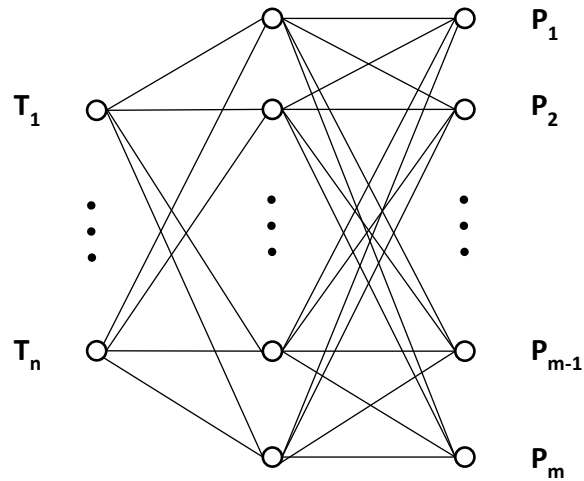

Figure 2: Neural Network Structure

The inputs to the ANN are the temperatures for each phase and the outputs are the species properties of interest (specific heat or enthalpy for each species). Given the simplicity of the task, a single hidden layer was used with a sigmoid activation. The width of the hidden layer was set to the width of the number of output properties. The output layer used a linear activation. Figure 2 shows the simple structure for the ML estimator. Training was performed by randomly sampling across the temperature range for the polynomials independently for each phase to generate species thermodynamic properties. The data was split into two groups for training and testing. Given the low number of free parameters in the model and ease of generating training data, the input domain was simply heavily over sampled to prevent over fitting. Optimization of the neural network was performed with a custom solver implementation of the Levenberg Marquardt Algorithm (LMA) in TensorFlow. The LMA is a second-order (requires an estimation of the Jacobian matrix) optimizer known to be much more efficient than first-order methods when the number of samples and variables is small.

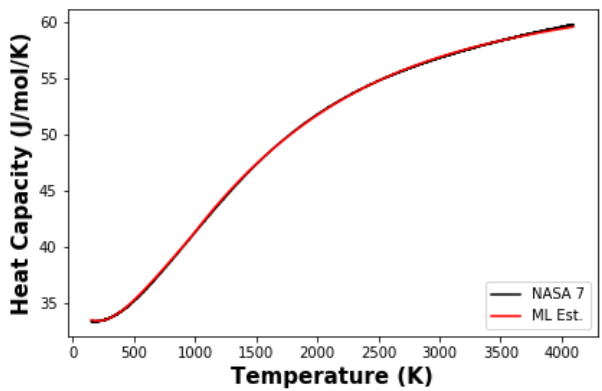

*Figure 3: Example ML model estimation for the specific heat of water vapor*

Figure 3 shows the ML estimation and training data for the specific heat of water vapor. The ANN has a maximum of 0.436% difference from the polynomial calculation and an average difference of 0.138%.

Even though the thermodynamic properties are traditionally evaluated via simple polynomials, it was found that replacing them with an ANN estimation saved considerable calculation time. On average, evaluating the ANN was about 20% faster than calculating all the species properties via the polynomials. The thermodynamic calculations needed to calculate the rate of change of temperature are not an insignificant part of each iteration. The exact amount depends on the number of species relative to the number of reactions and operations in each reaction, but most global mechanisms are roughly equal. Experience has shown that the thermodynamic calculations may account for roughly half of the computational effort for global mechanisms. Thus the 20% speed up obtained via ANN estimators translates to about a 10% reduction in time to solution. The speed gains could be even more substantial on a Volta architecture if the mixed precision tensor cores are used for the evaluation of the ANNs as they are known to accelerate ANN evaluations by as much as 1.8 times. When the sampling and training of the ANN is so quick and automated using the LMA with oversampling, this speed gain is simple and quick to obtain.

## 4. The Absolute Stability Analysis of the Forward Euler STEV Solver

Assuming that reactions are elementary, the rates of change of the thermochemical state are a function of the thermochemical state as in Eq (17).

$$\boldsymbol{X'} = \boldsymbol{f}(\boldsymbol{X}) \qquad (17)$$

The Forward Euler finite difference method is the simplest and most straightforward explicit solution estimator for ODEs. It is derived simply from the first order linear approximation of a differential equation by Taylor Series expansion resulting in the following scheme:

$$X_{i+1} = X_i + hf(X_i, t_i) \tag{18}$$

Where $x$ represents the state variable at a point in time $i$, $h$ is the step size between the current and the next point in time, and $f$ is the right-hand side function evaluated at the current step.

The absolute stability criteria for the Forward Euler method requires that the eigenvalues of the coefficient matrix be bounded by the disk of radius 1 centered at -1 in the left half of the complex plane:

$$|1 + h\lambda| \leq 1 \tag{19}$$

Where $\lambda$ is a representative eigenvalue. A first order approximation of the non-linear representative system of equations shown in Eq (17) by Taylor Expansion is:

$$X := X_0 + \Delta X \tag{20}$$

For small enough step changes $\Delta X$ in the state vector variable $X$ from an initial state $X_0$.

$$X_0' + \Delta X' \approx f(X_0) + \left.\frac{\partial f}{\partial X}\right|_{X_0} \Delta X \tag{21}$$

Since $X_0' = f(X_0)$, a new linearized approximation emerges as:

$$\Delta X' = \left.\frac{\partial f}{\partial X}\right|_{X_0} \Delta X \tag{22}$$

The local stability of Eq (22) is found by determining the eigenvalues of the Jacobian matrix evaluated at the initial condition as applied to Eq (19):

$$A := \left.\frac{\partial f}{\partial X}\right|_{X_0}, \rho(A) = sup\{|\lambda|: \lambda \in \sigma(A)\} \tag{23}$$

Where $\rho(A)$ is the spectral radius of the Jacobian matrix and $\lambda$ is an eigenvalue of the spectrum, $\sigma(A)$. The eigenvalues for a real chemically reactive system are all negative and real,[11] leading to the following stability criterion from Eq (19).

$$h \leq \frac{2}{\max_i \sum_{j=1}^n |a_{ij}|} = \frac{2}{\|A\|_\infty} \leq \frac{2}{\rho(A)} = \frac{2}{sup|\lambda|} \leq \frac{2}{|\lambda|} \ \forall \lambda \in \sigma(A) \tag{24}$$

Combining the stability criterion in Eq (24) with the time step determined from Eq (5) gives the absolute stability criterion for the STEV solver when the Euler Method is used.

$$h = min\left(\min_i \left(\frac{\min(Y_{step}^{max}, y_i^*)}{r_i^*}\right), \Delta t_{max}\right) \leq \frac{2}{\max_i \sum_{j=1}^n |a_{ij}|} \tag{25}$$

# 5. CFD, TensorFlow, and the STEV Solver

In most modern CFD codes, integration of stiff chemistry is done by operator splitting. That is, the fluid is evolved separately from the chemistry so that the fluid equations can be solved on a larger time step and the chemistry can be solved on a substantially smaller step which saves computational effort without a significant impact on accuracy. Without operator splitting, the entire system of coupled equations would have to be solved on at chemistry time step. As a demonstration, NETL's MFiX CFD code[1] was coupled to TensorFlow using multilanguage compilation of a C interface that can pass memory directly to the python TensorFlow library which can execute the graph containing the STEV solver. In this way it functions as a drop-in replacement for traditional CFD Stiff Solvers in most any CFD code.

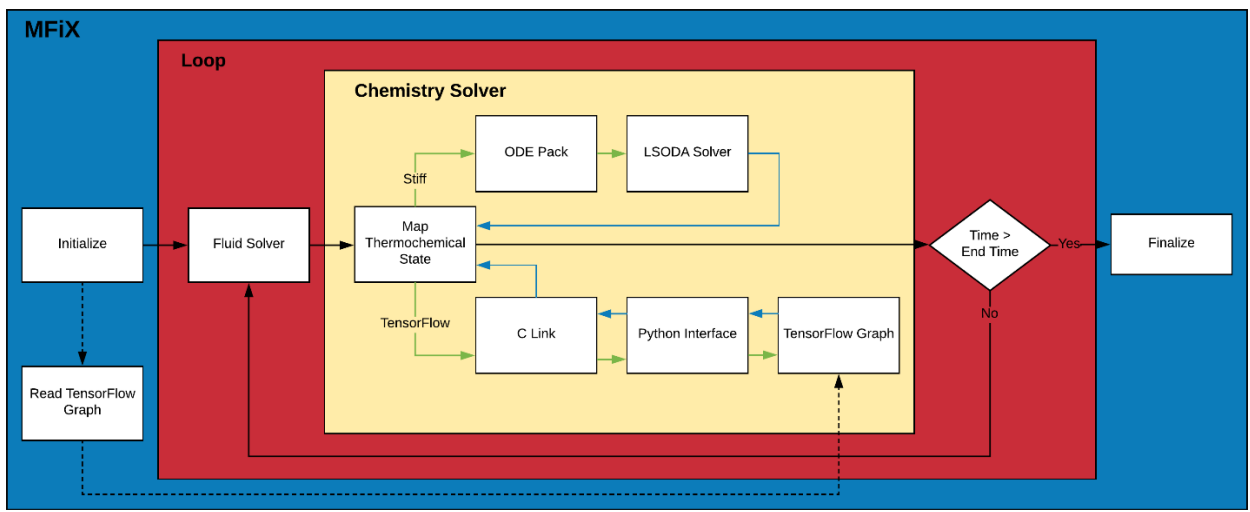

*Figure 4: Integration of the STEV solver through TensorFlow in MFiX*

To seamlessly integrate TensorFlow with MFiX, a switch statement was added enabling the user to choose to proceed through the original LSODA solver or bypass it through a simple variable declaration in the project file and perform the calculations within TensorFlow. An integration layer that consists of a C link and a Python interface was developed to provide bi-directional communication between MFiX and TensorFlow. The C link is called directly from the MFiX solver and provides the desired arguments in the form of memory pointers. Within the link, those pointers are then converted to python objects and passed along to the python interface. The interface then directly makes the necessary calls to the TensorFlow graph to perform the calculations. Those calculations are hardware agnostic and can be accelerated on any type of processing unit that TensorFlow supports. The results are then passed by reference back to the C link which in turn properly casts the memory and passes it back to the MFiX solver. MFiX then assigns those results to the thermochemical state. This implementation was chosen over the alternatives as the python API is the officially supported API for TensorFlow and it provides the most feature rich and simple interface. However, there are other ways to accomplish the same task which may be more efficient.

# 6. Computational Hardware

Unless otherwise specified, all computational work was completed on NETL's Joule 2.0 HPC. Joule 2.0 has a total of 78,720 cores, 242.304 TB of ram, and a peak performance of 5.17 PFlop/s. There are 1664 nodes. Each node has dual socket Xenon Gold 6148 with twenty cores for a total of forty CPU cores per node. One hundred nodes have two NVIDIA P100's each. Each simulation in this work was done on a single node. LSODA MFiX simulations were carried out in serial (on a single core) or by message passing interface (MPI) with 40 ranks on the same node. The STEV MFiX simulations were carried out through the TensorFlow framework which automatically utilizes all cores available on CPUs when executed on that device. The GPU implementation utilized a single P100 (future work will allow for multiple GPU utilization).

## 7. Results and Discussion

Two test cases were completed with the integrated MLA-STEV solver in MFiX. The first test case was a gas phase methane combustion case and the second was a traditional gasifier. Tests were completed using the implicit LSODA based CPU solver in MFIX, and this was compared to the MLA-STEV solver in TensorFlow based on the CPU and GPU.

### 7.1 Gas Phase Methane Combustion

Initial testing of the TensorFlow implemented STEV Solver was performed in a gas phase only case using the BFER mechanism for methane oxidation.[7] The BFER mechanism consists of a stiff methane partial oxidation reaction (MEPOX) and a stiff CO reversible oxidation (COROX) reaction and is summarized in Eqs (26) to (30) and Table 1.

$$CH_4 + 1.5\ O_2 \xrightarrow{fwd} CO + 2\ H_2O \quad (26)$$

$$CO + 0.5\ O_2 \xleftrightarrow{rev} CO_2 \quad (27)$$

$$r_{MEPOX} = k_{MEPOX} e^{-E_a/RT} C_{CH_4}^{0.5} C_{O_2}^{0.65} \quad (28)$$

$$r_{COROX} = k_{COROX} T^{0.7} e^{-E_a/RT} \left( C_{CO} C_{O_2}^{0.5} - \frac{C_{CO_2}}{K} \right) \quad (29)$$

$$K = \left( \frac{1\ atm}{RT} \right)^{\Delta N} e^{-\Delta G_{rxn}/RT} \quad (30)$$

*Table 1: BFER reaction mechanism constants*

| Reaction | k | $E_a\ (cal/mol)$ |
|---|---|---|
| MEPOX | 4.9E+09 | 35500 |
| COROX | 2.00E+08 | 12000 |

*Reaction set based on the following units: cm, s, mol, cal*

Here, $k$ is the rate constant, $E_a$ is activation energy, $R$ is the ideal gas constant, $T$ is temperature, $C$ is the concentration of the specie, $K$ is the equilibrium constant, $\Delta N$ is the change in moles based on reaction stoichiometry, and $\Delta G_{rxn}$ is the change in Gibbs free energy due to the reaction.

The first set of benchmakrs was conducted by running the solvers on the same set of random thermochemical states. Each chemistry solver was timed independently with an integration final time of 1 millisecond (a typical fluid time step) using increasing numbers of hypothetical cells. Since the thermochemical states are random but fixed, each solver operates on the same set of states at each fixed cell count. As the number of cells increases, the random states begin to cover the broad range of stiff and non-stiff combinations giving a more representative set of equations to solve. As can be seen from Figure 5, the time taken by the LSODA solver in MFiX exhibits an approximately linear and proportional behavior relative to the number of cells being solved. The MLA-STEV solvers run in TensorFlow exhibit a very different behavior with a nearly flat response out to a critical number of cells at which point memory saturation begins to affect the calculation speed. In as few as 200 cells, the CPU MLA-STEV solver implementation becomes faster than the LSODA solver, and in as few as 1000, the GPU MLA-STEV solver becomes faster. In large scale simulations, the GPU MLA-STEV solver has the capability of being over 200 times faster than the LSODA solver. It is also worth noting that even when memory saturated, the GPU implementation continues to grow in relative speed increase because the memory access rate on the GPU is substantially higher than the CPU.

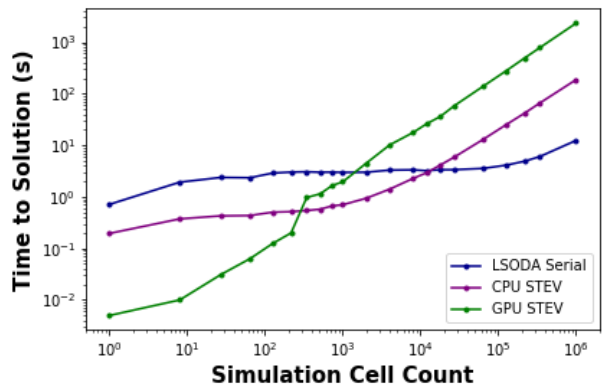

*Figure 5: Single phase chemistry with random cell compositions solver timing comparisons.*

To verify the implementation, a fully developed diffuse flame CFD simulation run in MFiX. The flame was established using the default LSODA MFiX Stiff Chemistry Solver and run for several seconds to eliminate the effects of the initial condition. The restart checkpoint was used as the base for running the simulation an additional ten seconds of simulation time using each of the default MFiX Stiff Chemistry Solver, MLA-STEV TensorFlow CPU solver, and MLA-STEV TensorFlow GPU solver. The final second of each simulation was time averaged. The domain was sliced vertically in the middle of the flame and each cell was compared between the default LSODA MFiX Stiff Chemistry solver and the CPU/GPU MLA-STEV Solver was calculated for each thermochemical property. The results from the CPU/GPU MLA-STEV solvers were virtually identical. Figure 7-Figure 9 illustrates the minute differences between the LSODA solver and the GPU MLA-STEV solver. The differences only occur along the reaction front. As can be seen in the figures, the simulations are virtually identical and match to within 0.006% in every case.

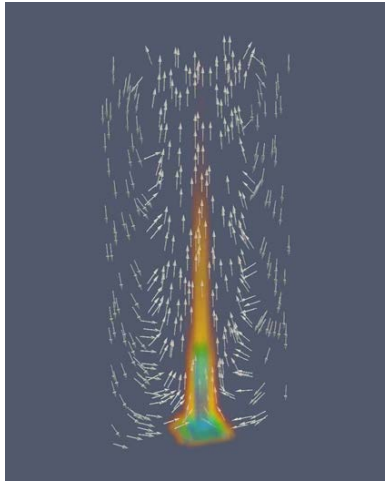

*Figure 6: Gas Phase Chemistry in Diffuse Flame CFD simulation. Colored by combustion species*

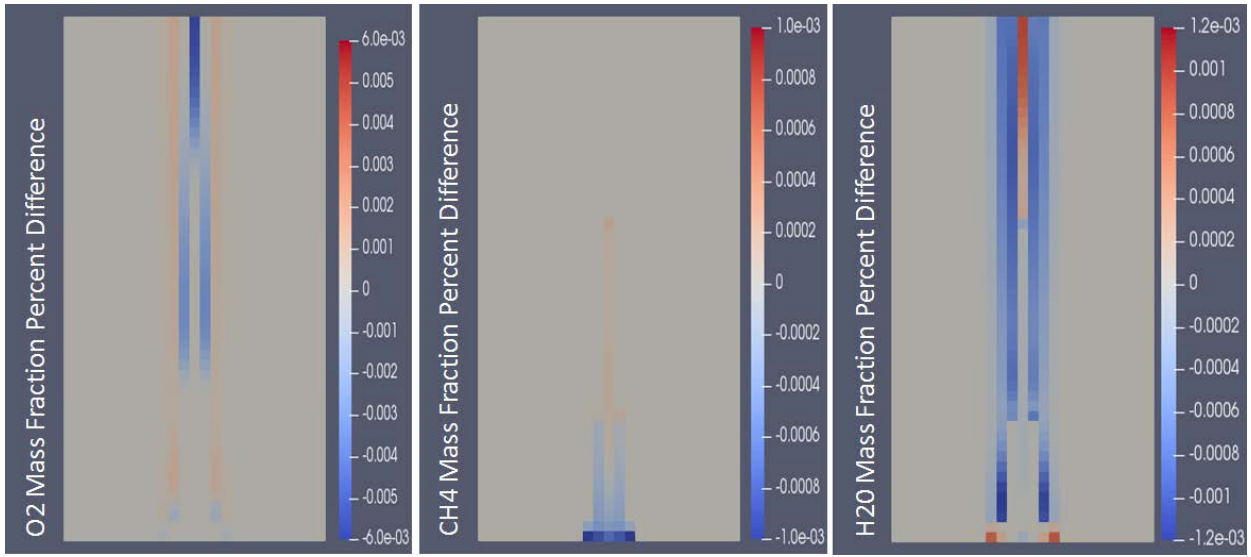

*Figure 7 - Percentage difference between the LSODA Solver and the MLA-STEV solver for O2, CH4, and H2O mass fractions respectively*

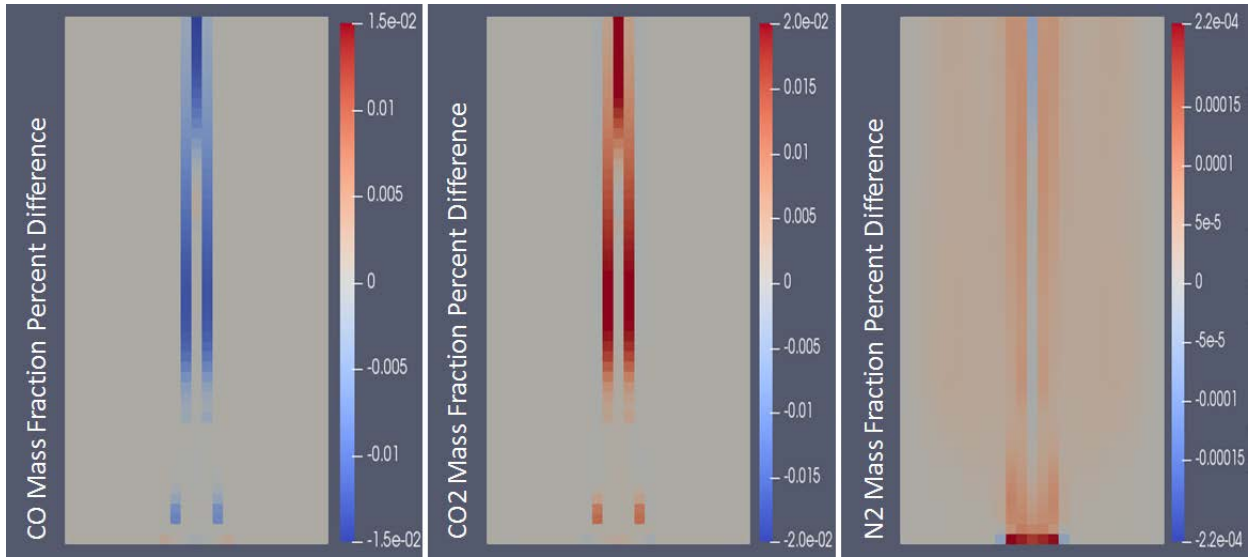

*Figure 8: Percentage difference between the LSODA Solver and the MLA-STEV solver for CO, CO2, and N2 mass fractions respectively*

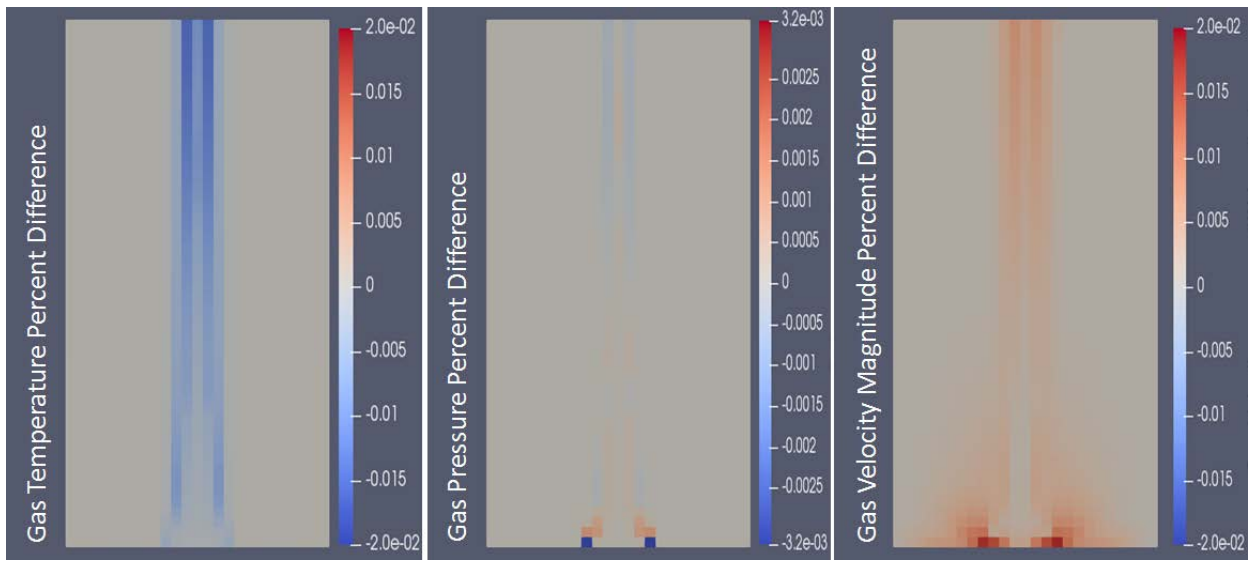

*Figure 9: Percentage difference between the LSODA Solver and the MLA-STEV solver for Temperature, Pressure, and Velocity Magnitude respectively*

## 7.2 Multiphase Coal Gasification

Based on the success of the single-phase combustion case, the STEV solver was extended to handle multiphase reactions. NETL's Carbonaceous Chemistry for Computational Modeling (C3M) was utilized to generate the chemical kinetics for gas-solid multiphase gasification as summarized in Eqs (31) to (56) and Table 2.[12]

$$CV(s) \xrightarrow{PYRO} 0.242057\, CO(g) + 0.0987274\, CO_2(g) + 0.229174\, CH_4(g) +$$
$$0.0852648\, H_2(g) + 0.198778\, H_2O(g) + 0.0112077\, H_2S(g) + 0.024509\, C_3H_6(g) +$$
$$0.0266777\, HCN(g) + 0.0473162\, C_2H_4(g) + 0.008257\, C_2H_6(g) + 0.02146\, Tar(g) \quad (31)$$

$$CO(g) + H_2O(g) \xleftrightarrow{WGS} CO_2(g) + H_2(g) \quad (32)$$

$$CC(s) + H_2O(g) \xrightarrow{STG} CO(g) + H_2(g) \quad (33)$$

$$CC(s) + CO_2(g) \xrightarrow{CG} 2CO(g) \quad (34)$$

$$CC(s) + 2H_2(g) \xrightarrow{MEG} CH_4(g) \quad (35)$$

$$CC(s) + O_2(g) \xrightarrow{COX} CO_2(g) \quad (36)$$

$$Tar(g) + 28.6409\, O_2(g) \xrightarrow{TOX} 25.1144\, CO_2(g) + 11.9133\, H_2O(g) + 0.0881788\, SO_3(g)$$
$$+ 0.543006\, N_2(g) \quad (37)$$

$$CO(g) + 0.5O_2(g) \xrightarrow{COOX} CO_2(g) \quad (38)$$

$$H_2(g) + 0.5O_2(g) \xrightarrow{HOX} H_2O(g) \quad (39)$$

$$CH_4(g) + 2O_2(g) \xrightarrow{MEOX} CO_2(g) + 2H_2O(g) \quad (40)$$

$$H_2O(s) \xrightarrow{MR} H_2O(g) \quad (41)$$

$$r_{PYRO} = A_{PYRO}\varepsilon_s\rho_s X_{CV}\gamma(X_{CV})e^{-Ea_{PYRO}/T_s} \quad (42)$$

$$r_{WGS} = \varepsilon_g\varepsilon_s\rho_s(A_{WGS}P)^{(-F_{P,WGS}P + 0.5)}\left(K_{F,WGS}X_{CO}X_{H_2O}\gamma(X_{CO})\gamma(X_{H_2O})e^{-Ea_{WGS}^F/T_g}\right.$$
$$\left. - K_{R,WGS}X_{CO_2}X_{H_2}\gamma(X_{CO_2})\gamma(X_{H_2})e^{-Ea_{WGS}^R/T_g}\right) \quad (43)$$

The gasification reactions ($r_{STG}, r_{CG}, r_{MEG}$) and char oxidation ($r_{COX}$) were derived from PC Coal Lab (PCCL) in C3M. The reaction rates from PCCL utilize an annealing factor, $\alpha_i$, which is based on the reaction progress of conversion of fuel to ash. To keep the system stable in CFD, Eq (44) calculates the reaction progress factor, $\vartheta$. The min and max statements help keep the reaction progress variable between zero and unity if the mass fractions are zero or negative. Eqs (47) to (50) use the annealing factor found in Eqs (44) and (46). The i in Eq (46) is relative to gasification denoted by "G" or char combustion denoted by "C".

$$\vartheta = max\left(0, min\left(1, \beta_c \frac{X_{CC}}{X_{CA}}\right)\right) \tag{44}$$

$$\varphi = 1 - \vartheta \tag{45}$$

$$\alpha_i = \varphi^5 + \theta_4^i \varphi^4 + \theta_3^i \varphi^3 + \theta_2^i \varphi^2 + \theta_1^i \vartheta + \theta_0^i \tag{46}$$

$$r_{STG} = A_{STG} \alpha_G \varepsilon_s \rho_s X_{CC} \gamma(X_{CC}) \frac{\left(\frac{MW_g P_g X_{H_2O}}{18.01530}\right)^{0.98}}{1 + F_{P,STG}\left(\frac{MW_g P_g X_{H_2}}{2.01588}\right)} e^{-Ea_{STG}/T_s} \tag{47}$$

$$r_{CG} = A_{CG} \alpha_G \varepsilon_s \rho_s X_{CC} \gamma(X_{CC}) \frac{\left(\frac{MW_g P_g X_{CO_2}}{44.00980}\right)^{0.56}}{1 + F_{P,CG}\left(\frac{MW_g P_g X_{CO}}{28.01040}\right)} e^{-Ea_{CG}/T_s} \tag{48}$$

$$r_{MEG} = A_{MEG} \alpha_G \varepsilon_s \rho_s X_{CC} \gamma(X_{CC}) \left(\frac{MW_g P_g X_{H_2}}{2.01588}\right) e^{-Ea_{MEG}/T_s} \tag{49}$$

$$r_{COX} = \alpha_C A_{COX} \gamma(X_{O_2}) \gamma(X_{CC}) e^{-Ea_{COX}/T_{film}} \left(\frac{MW_g P_g X_{O_2}}{31.9988}\right)^{0.73} \varepsilon_s/d_p \tag{50}$$

$$T_{film} = 0.5T_g + 0.5T_s \tag{51}$$

$$r_{TOX} = A_{TOX} e^{-Ea_{TOX}/T_g} \varepsilon_g \rho_g^{1.75} X_{O_2}^{1.5} X_{Tar}^{0.25} \gamma(X_{O_2}) \gamma(X_{Tar}) \tag{52}$$

$$r_{COOX} = A_{COOX} e^{-Ea_{COOX}/T_g} \varepsilon_g \rho_g^{1.75} X_{O_2}^{0.25} X_{CO} X_{H_2O}^{0.5} \gamma(X_{O_2}) \gamma(X_{CO}) \tag{53}$$

$$r_{HOX} = A_{HOX} e^{-Ea_{HOX}/T_g} \varepsilon_g \rho_g^2 X_{O_2} X_{H_2} \gamma(X_{O_2}) \gamma(X_{H_2}) \tag{54}$$

$$r_{MEOX} = A_{MEOX} e^{-Ea_{MEOX}/T_g} \varepsilon_g \rho_g^{1.5} X_{O_2}^{1.3} X_{CH_4}^{0.2} \gamma(X_{O_2}) \gamma(X_{CH_4}) \tag{55}$$

$$r_{MR} = A_{MR} \gamma(X_{CM}) \varepsilon_s / d_p \tag{56}$$

Here, $X$ is the mass fraction and the subscript denote that species formula. The species formula is symbolic of the most common compounds represented by those formulas. For example, $O_2$ is molecular oxygen and $CH_4$ is methane. However, $CV$ is the coal volatile matter, $CC$ is the coal fixed carbon, $CM$ is the coal moisture, $CA$ is the coal ash, and $Tar$ is the tar produced by pyrolysis as a complex mixture of hydrocarbons. $A$ is the pre-exponential factor and $Ea$ is the activation energy. The subscript denotes the reaction they are a part of. There are also several phase properties denoted by a subscript g for gas or subscript s for solid. $T$ is the phase temperature, $\varepsilon$ represents the phase volume fraction, $\rho$ is the phase density, $MW$ is the mixture molecular weight, and $P$ is the pressure. $d_p$ is particle diameter and $\gamma$ is the rate limit function defined in Eq (7). The constants for the model are defined in Table 2.

*Table 2: Multiphase Model Constants*

| Variable | Value |
|---|---|
| $A_{PYRO}$ | $5.39808917197452 \times 10^{-1}$ |
| $Ea_{PYRO}$ | $2.02305749338465 \times 10^{3}$ |
| $A_{WGS}$ | $9.86923266716013 \times 10^{-6}$ |
| $F_{P,WGS}$ | $3.94769306686405 \times 10^{-8}$ |
| $K_{F,WGS}$ | $2.87480073925354 \times 10^{0}$ |
| $Ea_{WGS}^{F}$ | $8.42570940582151 \times 10^{3}$ |
| $K_{R,WGS}$ | $1.08482987533646 \times 10^{2}$ |
| $Ea_{WGS}^{R}$ | $1.23814194058215 \times 10^{4}$ |
| $\theta_4^G$ | $-1.74119074065241 \times 10^{0}$ |
| $\theta_3^G$ | $6.75755882493237 \times 10^{-1}$ |
| $\theta_2^G$ | $2.42994500814316 \times 10^{-1}$ |
| $\theta_1^G$ | $1.80861879808380 \times 10^{-1}$ |
| $\theta_0^G$ | $-1.37554327457087 \times 10^{-1}$ |
| $\theta_4^C$ | $-1.71305029851856 \times 10^{0}$ |
| $\theta_3^C$ | $1.02332704517333 \times 10^{0}$ |
| $\theta_2^C$ | $-2.12046897502992 \times 10^{-1}$ |
| $\theta_1^C$ | $-4.34614630452274 \times 10^{-3}$ |
| $\theta_0^C$ | $1.00829969505805 \times 10^{-2}$ |
| $A_{STG}$ | $3.64422300238124 \times 10^{-5}$ |
| $F_{P,STG}$ | $2.58573895879595 \times 10^{-7}$ |
| $Ea_{STG}$ | $4.27760404137599 \times 10^{3}$ |
| $A_{CG}$ | $1.21674169495915 \times 10^{1}$ |
| $F_{P,CG}$ | $2.57586972612879 \times 10^{-6}$ |
| $Ea_{CG}$ | $1.76136636997835 \times 10^{4}$ |
| $A_{MEG}$ | $1.66401967700346 \times 10^{-8}$ |
| $Ea_{MEG}$ | $9.15910512388742 \times 10^{3}$ |
| $\beta_c$ | $4.7741935483871 \times 10^{-1}$ |
| $A_{COX}$ | $2.16444452624815 \times 10^{1}$ |
| $Ea_{COX}$ | $1.43928313687756 \times 10^{4}$ |
| $A_{TOX}$ | $2.59900518618807 \times 10^{6}$ |
| $Ea_{TOX}$ | $1.50981378963261 \times 10^{4}$ |
| $A_{COOX}$ | $7.91515800843217 \times 10^9$ |
| $Ea_{COOX}$ | $2.01288244766506 \times 10^4$ |
| $A_{HOX}$ | $1.67426958313306 \times 10^{11}$ |
| $Ea_{HOX}$ | $1.50966183574879 \times 10^4$ |
| $A_{MEOX}$ | $1.34383699737932 \times 10^9$ |
| $Ea_{MEOX}$ | $2.43558776167472 \times 10^4$ |
| $A_{MR}$ | $1.0061447769396 \times 10^{-1}$ |

Like the single-phase benchmarks, the multiphase chemistry was tested with a set of randomized thermochemical states used in each implementation. As in the single-phase case, the time to solution for the LSODA MFiX Stiff Chemistry solver increased linearly and proportionally to the number of cells being solved and there was also a similar trend for the MLA-STEV solvers in TensorFlow. Again, the CPU and GPU MLA-STEV solvers became faster than the LSODA solvers at roughly 100 and 1000 cells respectively. From Figure 10 the crossover point in solution time for the number of cells the CPU and GPU implementation of the STEV solver has extended to a much higher cell count versus the single phase example. This is likely due in part to the increased communication overhead needed to transfer a much larger thermochemical state on and off the GPU. For the single-phase case, only 8 thermochemical variables per cell needed to be transferred. For the multiphase case, there were 26 variables per cell. Even though there were lesser speed gains for the GPU STEV solver relative to the CPU STEV solver, the gains over the LSODA solver were substantial.

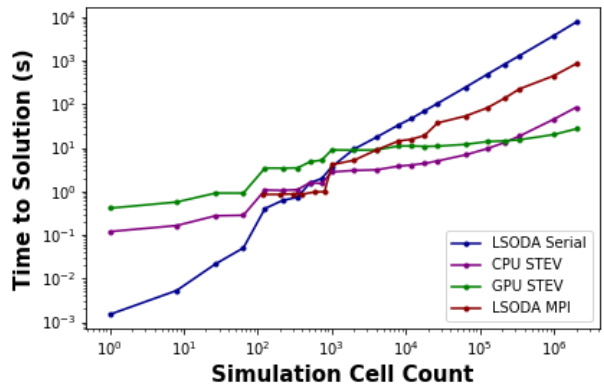

Figure 10: Multiphase chemistry with random cell compositions solver timing comparisons.

## 8. TRIG Gasifier Comparisons

To further benchmark and verify the implementation several simulations of the Power Systems Development Facility (PSDF) in Wilsonville Alabama were conducted. The PSDF is a pilot scale transport gasifier operated by Southern Company and co-funded by the United States Department of Energy. This system was used to develop the Transport Reactor Integrated Gasification (TRIG[TM]) technology for clean coal power and chemical systems. The TRIG[TM] reactor is a circulating fluidized bed reactor like the fluidized catalytic cracking reactors developed for the oil and gas industries. The system consists of two long vertical pipes in which solids circulate, a riser and a standpipe. Gas and a small amount of solids are injected into the bottom of the riser. The gas velocity is high enough to carry the solids upwards (ie transport the solids). At the top, set of cyclones which separates solids and gas. The solids drop down the standpipe to a j-leg and are injected back into the bottom of the riser. The whole system is operated at elevated temperatures but below the ash fusion temperature. Coal is injected into the riser at various points and is gasified to produce a syngas mixture. The system was developed to gasify low rank coals. In these tests, only the riser was simulated. Figure 11 shows a schematic representation of the TRIG[TM] system by Kellogg Brown & Root, LLC (KBR).

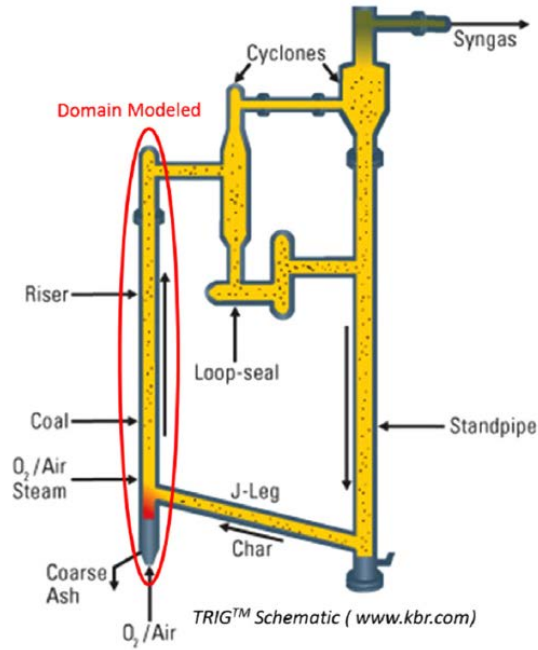

Figure 11: Schematic Representation of the TRIG system, by permission[13]

To further understand the speed gains of the STEV solvers in the context of real world use, a second set of simulations was completed in which the number of cells of a gasifier were varied using the same setup from past analytical work at NETL with the PSDF.[13] The chemistry models were replace with the above chemistry model from C3M, and the grid resolution was varied.

To spin up the simulations at each resolution up, the GPU MLA-STEV solver was used to simulate twelve seconds of run time. The residence time was approximately two seconds which allowed for more than six reactor volume changes to wash out the initial condition. Following this period, each simulation was benchmarked using each chemistry solver for two seconds using the same restart file at 12 seconds. Following that, the GPU STEV solver and stiff solver were restarted and run for an additional twenty seconds. The last ten were used as a comparison for solution accuracy. Figure 12 shows a typical averaged solids concentration path to pseudo steady state. The verification was done at near steady state.

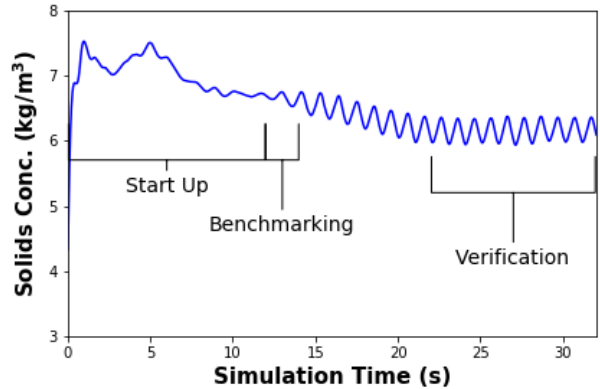

*Figure 12 - Simulation time evolution of analysis*

The left image in Figure 13 shows the time to solution for the GPU STEV solver, CPU STEV solver, and LSODA solver in serial and MPI as a function of grid resolutions for the two second benchmarking runs. The domain was decomposed for the MPI runs by slicing the gasifier into 40 equal pieces along the vertical axis of the gasifier. The right image shows both the fluid solver time (in dark colors) and the chemistry solver as a light shaded region.

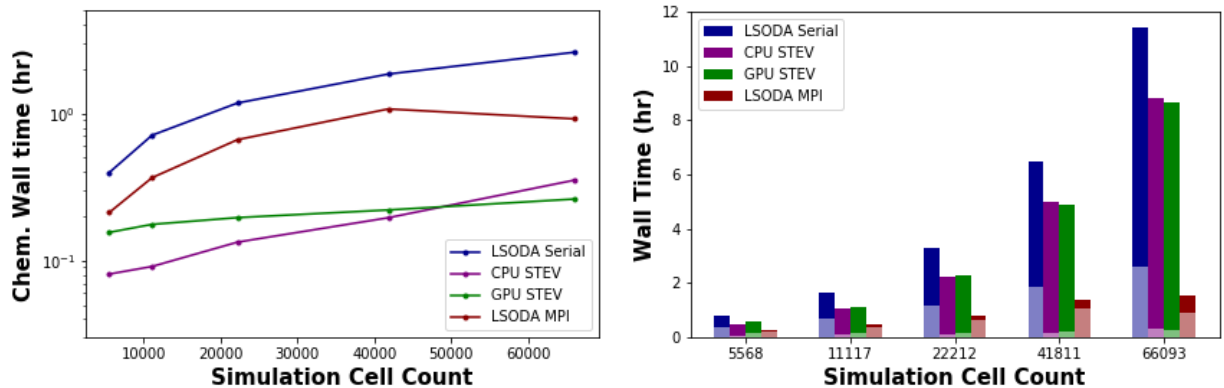

*Figure 13 – Total chemistry solve time (light region) as a fraction of total simulation time*

The time to solution was dramatically reduced using the STEV solver. On average, the STEV solver cut the chemistry solve time by an order of magnitude relative to the serial LSODA case. The MPI methods were substantially effective in cutting the fluid solver time but reduced the chemistry solution time by less than a factor of two. The relatively poor performance for the MPI for chemistry implementation is due to characteristically poor load balancing. The poor load balancing occurs because stiff cells are unevenly distributed in the domain and generally exist where hot gas is filled with both oxidant and fuel. Since the conditions are transient, efficiently decomposing the domain is difficult and often contradicts the decomposition needs of other load balancing needs. The STEV solver takes the entire domain and solves in parallel on a single device. Furthermore, the GPU implementation scales exceptionally well as problem size grows suggesting its appropriate use for large problem sets. While it is tempting to develop an MPI compatible version of the STEV solver, research direction is indicating that it would be more beneficial to implement a TensorFlow compatible fluid solver and link the existing TensorFlow compatible STEV solver with the new fluid code to take advantage of the GPUs and new hardware being developed.

The following shows numerical output data of two simulations of the PSDF TRIG reactor. The first simulation was done with the standard MFiX software with the default LDODA tolerances (rtol = 1x10$^{-5}$, atoll = 1x10$^{-6}$). The second is from the GPU enabled MLA-STEV solver.

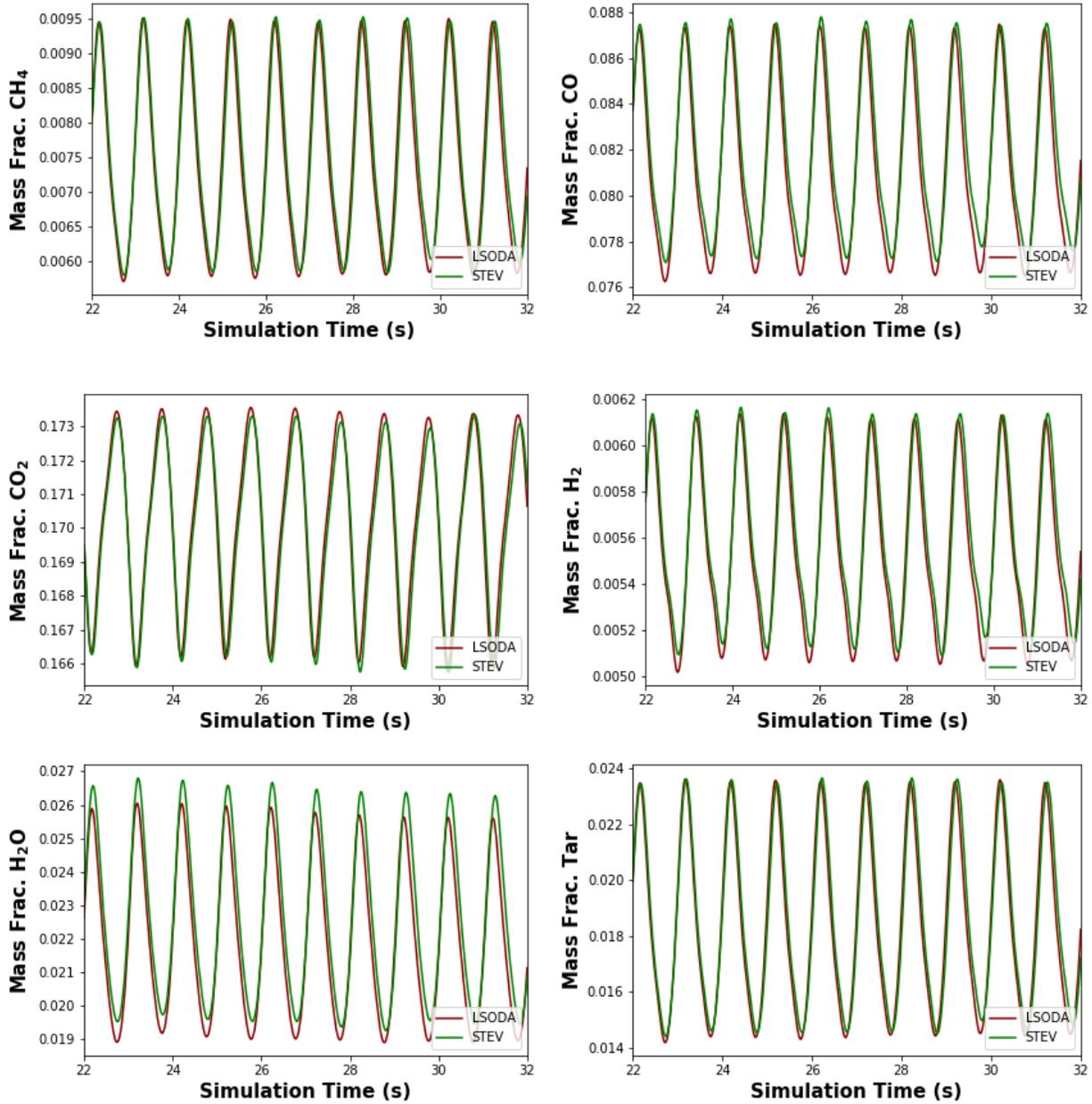

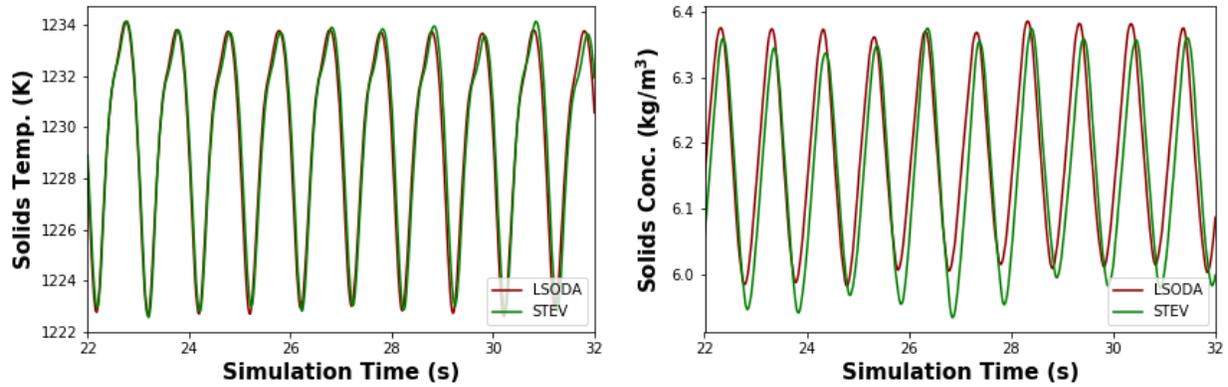

Figure 14: Thermochemical Properties as a function of simulation time

Figure 14 shows the mass fractions of the major gas species and solids temperate at the exit of the gasifier and the volume averaged solids concentration of the entire reactor as a function of the last ten seconds of simulation time. Given the high degree of complexity of the transient, reacting, multiphase simulations and the elapsed time from restart for each simulation, the simulations exhibit substantial agreement which indicates that the MLA-STEV solver is an appropriate technique for solving multiphase chemistry problems. Table 3 shows the mean, standard deviation and relative error for the properties plotted in Figure 14. Every property matched within 1% relative error, except for steam mass fraction which matched within 2.81%. While these results are quite close, the differences may be due to solving the thermochemical state with 32 bit precision and the associated value truncation when transitioning back to 64 bit for the rest of the fluid solver. However, determining their exact cause is difficult as the simulations are coupled sets of a large number of differential equations. Over time, numerical noise propagates and disturbs the solution pathway.

Table 3 - Temporal Statistics for the LSODA and STEV solver Implementations

| Property | Solver | Mean | St. Dev. | Relative Err |
|---|---|---|---|---|
| CH4 | LSODA | 0.007414 | 0.001260172 | 0.9384% |
| CH4 | STEV | 0.007483 | 0.00124104 | 0.9384% |
| CO | LSODA | 0.081229 | 0.003652789 | 0.7050% |
| CO | STEV | 0.081802 | 0.003490606 | 0.7050% |
| CO2 | LSODA | 0.170155 | 0.002473718 | 0.1072% |
| CO2 | STEV | 0.169973 | 0.002430991 | 0.1072% |
| H2 | LSODA | 0.005544 | 0.000354283 | 0.7893% |
| H2 | STEV | 0.005588 | 0.000342877 | 0.7893% |
| H2O | LSODA | 0.021983 | 0.002362653 | 2.8088% |
| H2O | STEV | 0.022601 | 0.002431039 | 2.8088% |
| TAR | LSODA | 0.018421 | 0.003131313 | 0.9391% |
| TAR | STEV | 0.018594 | 0.003083801 | 0.9391% |
| Temp. | LSODA | 1229.384629 | 3.726423188 | 0.0076% |
| Temp. | STEV | 1229.47784 | 3.690366521 | 0.0076% |
| | LSODA | 6.186609384 | 0.127418721 | 0.4605% |
| Solids Conc. | STEV | 6.158122304 | 0.134606772 |  |

## 9. Conclusions

The STEV solver was able to quickly and accurately predict the thermochemical state and dramatically reduce total simulation time through massive and efficient parallelization.  The STEV solver was shown to be able to solve very stiff chemical equations much faster than the state-of-the-art implicit methods if there were few stiff, reversible, *and* competitive reactions present.  The MLA-STEV solver was shown to be approximately 10% faster than the STEV solver and could be as high as 28% with ML specific accelerators.  The GPU version of the STEV solver was found to be as much as 200 times faster than the LSODA CPU implementation.  The CPU STEV solver tends to be faster than the GPU STEV solver when the cell count is lower and when mechanism size is large.  However, for industrially relevant simulation sizes of multiple millions of cells, the GPU is likely to always be preferable.

Some very important lessons were learned in the development of the STEV and MLA-STEV solvers.  First, even simple ML methods can have a profound impact on the workload when hybridized into traditional CFD algorithms.  Second, extreme care should be taken when hybridizing algorithms that work on primary state variables in CFD and it is probably best to limit use to a predictor/corrector scenario where ML is used to produce a smart guess and traditional algorithms are used to correct and enforce convergence.  Third, the nature of the solution methods for IVPs makes using ML exceptionally difficult if used on primary state variables because it is difficult to correct problems, but ML is suitable for secondary property prediction.  The nature of boundary value problems makes them much easier to hybridize as the field can be predicted by ML and corrected to a defined tolerance more easily.  As such the focus of this effort has shifted to accelerating linear and non-linear solvers in CFD and away from further acceleration of stiff chemistry.

## 10. Future Work

Because of the ease of defining device work using TensorFlow, the STEV solver could be further accelerated by exploring methods to divide the workload across multiple devices.  It should be possible to avoid memory saturation for the same number of cells through this methodology which could greatly increase the speed for large simulation.

The parameters used in controlling the size of the time step can be optimized to result in a higher allowable time step while remaining stable and respecting potential issues with delayed ignition seen due to overstepping and under stepping from the Euler Method.

Different methods for controlling reaction rates near equilibrium could be explored to improve performance when many stiff, reversible, and competitive reactions are present.  Perhaps ML methodologies could be employed to improve detection and response.

## 11. Acknowledgements

## 12. Disclaimer